\documentclass[aps,prl,notitlepage,%
superscriptaddress,longbibliography,twocolumn]{revtex4-1}

\usepackage[colorlinks=true, allcolors=blue]{hyperref}
\usepackage{natbib}
\usepackage{graphicx}
\usepackage{amssymb}
\usepackage{amsmath}
\usepackage{xcolor}
\usepackage{bm}
\usepackage{bbm}
\usepackage{dcolumn}
\usepackage{siunitx}
\usepackage[T1]{fontenc}

\allowdisplaybreaks

\newcolumntype{d}[1]{D{.}{.}{#1}}
\newcolumntype{.}{D{x}{}{9}}
\newcolumntype{,}{D{x}{}{5}}
\newcolumntype{;}{D{x}{}{19}}

\newcommand{\M}{\mathcal M}

\newcommand{\ep}{\epsilon}

\newcommand{\be}{\begin{equation}}
\newcommand{\ee}{\end{equation}}
\newcommand{\bearray}{\begin{eqnarray}}
\newcommand{\eearray}{\end{eqnarray}}
\newcommand{\bea}{\begin{eqnarray}}
\newcommand{\eea}{\end{eqnarray}}
\newcommand{\bse}{\begin{subequations}}
\newcommand{\ese}{\end{subequations}}
\newcommand{\bcases}{\begin{cases}}
\newcommand{\ecases}{\end{cases}}
\newcommand{\bpm}{\begin{pmatrix}}
\newcommand{\epm}{\end{pmatrix}}

\def\dbar{{\mathchar'26\mkern-11mud}}
\newcommand{\crr}{\nonumber \\}

\newcommand{\half}{\frac{1}{2}}

\definecolor{garrosgreen}{rgb}{0.1, 0.7, 0.1}
\definecolor{dartmouthgreen}{rgb}{0.05, 0.5, 0.06}
\definecolor{duelferred}{rgb}{0.7, 0.2, 0.1}
\definecolor{cambridgeblue}{rgb}{0.1, 0.3, 1.0}
\definecolor{oxfordblue}{rgb}{0.05, 0.2, 0.7}

\begin{document}

\title{Radiative-recoil corrections at order $\alpha^2 (Z \alpha)^5$ to energy levels of muonium and positronium}

\author{Gregory S. Adkins}
\affiliation{Department of Physics and Astronomy, Franklin \& Marshall College,
Lancaster, Pennsylvania 17604, USA}

\author{Addison Kovats-Bernat}
\altaffiliation[Present address: ]{Department of Physics, William \& Mary, Williamsburg, Virginia 23185, USA}
\affiliation{Department of Physics and Astronomy, Franklin \& Marshall College,
Lancaster, Pennsylvania 17604, USA}

\author{Elias Mitchell}
\affiliation{Department of Physics and Astronomy, Franklin \& Marshall College,
Lancaster, Pennsylvania 17604, USA}

\author{A. Volkan Turan}
\affiliation{Department of Physics and Astronomy, Franklin \& Marshall College,
Lancaster, Pennsylvania 17604, USA}

\begin{abstract}
We have calculated radiative-recoil corrections at order $\alpha^2 (Z \alpha)^5$ to the energy levels of spin-$\half$--spin-$\half$ Coulombic bound systems.  The corrections we consider involve two exchange photons with two-loop radiative corrections on the electron line.   We obtain precise results for all values of the mass ratio.  Our results are relevant for upcoming improved measurements of the muonium ground-state hyperfine splitting and 1S-2S interval.
\end{abstract}

\maketitle

{\em Introduction.---} Muonium and positronium are purely leptonic atoms, formed from an electron bound to either a positive muon or a positron.  The structure of these atoms is described almost entirely by QED, with only small contributions from strong and weak interaction effects.  Consequently, their energy levels can be calculated to high precision.  Corresponding high precision measurements allow for tests of current theory and improved determination of fundamental constants.  Representative reviews are given in Refs.~\cite{Sapirstein1990,Karshenboim2004,Karshenboim2005,Eides2007,Jungmann2016,Cassidy2018,Frugiuele2019,Adkins2022,Bass2023,Blumer2025}.  Novel approaches and improved and prospective measurements of the muonium \cite{Delaunay2021,Ohayon2022,Janka2022,Iwai2023,Cortinovis2023,Javary2025,Strasser2025,Torii2026,Blumer2025,Blumer2026} and positronium \cite{Sheldon2023a,Sheldon2023b,Gloggler2024,Shu2024,Heiss2025,Borges2026,Newson2026,Newson2026b} energy levels provide the incentive to complete the order $\alpha^7$ energy level corrections.

Of immediate importance is the completion of the order $\alpha^7$ corrections to the muonium ground-state hyperfine splitting.  New measurements of this splitting by the MuSEUM collaboration are currently underway at the J-PARC Muon Science Facility (MUSE), with a precision goal of $\sim \! 10 \, {\rm Hz}$ \cite{Iwai2023,Strasser2025,Torii2026}.  Current theory for this transition is \cite{Eides2007,Eides2019,Mohr2025,Eides2026}
\be
\Delta E_{\rm hfs}^{\rm th}({\rm Mu}) = 4\,463\,302\,873 (511)(70)(2) \, {\rm Hz} \, ,
\ee
where the quoted uncertainties are due to the uncertainty in the muon mass, uncalculated theoretical corrections, and the uncertainty in the fine structure constant $\alpha$.  The theoretical uncertainties are dominated by pure recoil corrections of order $m_r (Z \alpha)^7$ and radiative-recoil corrections of orders $m_r \alpha (Z \alpha)^6$ and $m_r \alpha^2 (Z \alpha)^5$, where $m_r$ is the muonium reduced mass and $Z$ is an assumed charge of the $\mu^+$ in units of the proton charge (used to keep track of the number of muon-photon interactions: for muonium, $Z=1$).  Many of the order $m_r \alpha^2 (Z \alpha)^5$ recoil corrections are already known \cite{Karshenboim1989,Eides2016,Eides2022}.  We have completed the remaining recoil correction of this order, involving two-photon exchange with two-loop radiative corrections on the electron line, with negligible uncertainty.

We have calculated the radiative-recoil corrections to muonium and positronium S-state energy levels coming from two-photon exchange with two-loop radiative corrections on the electron line as a function of the mass ratio $x = m_1/m_2$ between constituent particles.  Prior calculations \cite{Eides1993,Eides1994a,Kinoshita1994,Pachucki1994,Eides1994b,Eides1995b,Eides1995,Kinoshita1996,Mondejar2010,Dowling2010} took into account the reduced mass dependence of the wave function at zero distance $\vert \psi(0) \vert^2 = \frac{(m_r Z \alpha)^3}{\pi n^3}$, but otherwise, only the no-recoil limit $x=0$ was considered.  We have found high-precision results for systems with $x$ ranging from small (hydrogen, muonium, muonic hydrogen) to the maximum value $x=1$ as appropriate for positronium and true muonium.  

\begin{figure}[t!]
\begin{center}
\begin{minipage}{0.99\linewidth}
\begin{center}
\includegraphics[width=0.99\linewidth]{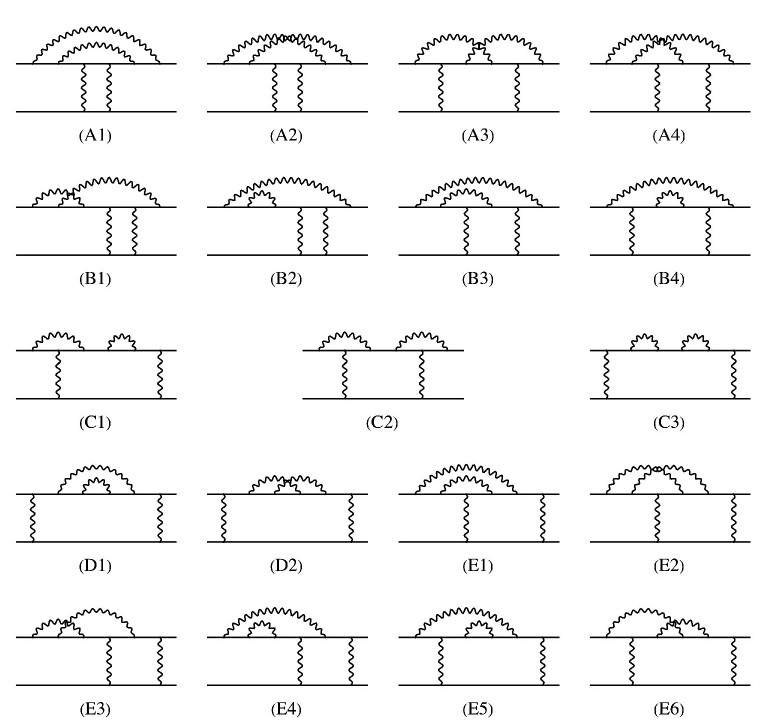}
\end{center}
\caption{\label{fig1}
The nineteen Feynman diagrams having no fermion loops that contribute to bound state energy levels at order $\alpha^2 (Z \alpha)^5$.  Also contributing, but not shown, are the nineteen graphs where the two exchange photons cross one another.  Some of the graphs [(A4), (B1), (B2), (B3), (C1) and group (E)] must be doubled to account for reflection symmetry.}
\end{minipage}
\end{center}
\end{figure}

{\em General Outline.---}
There are six sets of graphs that contribute to energy levels at order $\alpha^2 (Z \alpha)^5$, as displayed in Ref.~\cite{Karshenboim1989}.  Five of these sets involved vacuum polarization or light-by-light fermion loops, while the sixth contains two-loop radiative corrections to the electron line with no vacuum polarization contributions.  The five sets of graphs containing fermion loops have been studied in detail, including recoil corrections \cite{Eides2015,Eides2016,Eides2022}.  The last set of graphs is the one of our concern here.  The graphs of interest are shown in Fig.~\ref{fig1}.

Our calculations were done in the framework of NRQED \cite{Caswell1986}.  We have effectively calculated the $d$ matching coefficients of Eq.~(1) of \cite{Caswell1986} or Eq.~(22) of \cite{Kinoshita1996} that represent four-fermion contact terms.  The corresponding energy corrections are given by $\Delta E = - \vert \psi(0) \vert^2 \M$ where $\M$ is the order $\alpha^2 (Z \alpha)^2$ three-loop amplitude for scattering of the constituent particles at zero relative velocity ({\it i.e.} at threshold).

For illustration of the calculation, we consider the diagram shown in Fig.~\ref{fig1}(A1).  We calculate in $d=4-2\ep$ dimensions in order to regularize ultraviolet (UV) and infrared (IR) divergences.  We use a timelike metric ($g_{00}=+1$, $g_{i j} = -\delta_{i j}$), natural units $\hbar = c = 1$, and values of fundamental constants from \cite{Mohr2025}.  The amplitude $\M_{A1}$, coming from  the diagram of Fig.~\ref{fig1}(A1) is given by
\begin{widetext}
\bea
{\cal M}_{\rm A1} &=& -i \int \dbar^d \! q \, \dbar^d \! r \, \dbar^d \! s \, \overline{u}_{1 f} \left (-i q_1 \gamma^\nu \right ) \frac{i}{\gamma(m_1 n-q)-m_1} \left ( -i q_1 \gamma^\kappa \right ) \frac{i}{\gamma(m_1 n-q-r)-m_1} \left (-i q_1 \gamma^{\mu_1} \right ) \crr
&\hbox{}& \hspace{-1.8cm} \times \frac{i}{\gamma(m_1 n-q-r+s)-m_1} \left (-i q_1 \gamma^{\mu_2} \right ) \frac{i}{\gamma(m_1 n-q-r)-m_1} \left (-i q_1 \gamma_\kappa \right ) \frac{i}{\gamma(m_1 n-q)-m_1} \left (-i q_1 \gamma_{\nu} \right )u_{1i} \crr
&\hbox{}& \hspace{-1.8cm} \times \left ( \frac{-i}{q^2} \right ) \left ( \frac{-i}{r^2} \right ) \left ( \frac{-i}{s^2} \right )^2 \,
\overline{u}_{2 f}  \left (-i q_2 \gamma_{\mu_1} \right ) \frac{i}{\gamma(m_2 n-s)-m_2} \left (-i q_2 \gamma_{\mu_2} \right ) u_{2 i} \, ,
\eea
where $q_k$ and $m_k$ are the charge and mass of the $k^{\rm th}$ constituent, $n=(1,\vec 0\,)$ is a timelike unit vector, $m_k n$ is the incoming and outgoing energy-momentum vector of the $k^{\rm th}$ particle at threshold in the center-of-mass (CM) frame, and $\dbar^d \!q \equiv d^d \!q/(2 \pi)^d$.  We have used $q$ and $r$ for the momenta of the radiative correction photons and $s$ for the momentum of the exchange photons, and $u_{kf}$ and $u_{ki}$ are the final and initial Dirac spinors of the $k^{\rm th}$ constituent.  This amplitude can be written as
\bea
{\cal M}_{\rm A1} &=& i (q_1^2)^2 (q_1 q_2)^2 \int \dbar^d \! q \, \dbar^d \! r \, \dbar^d \! s \, \frac{\overline{u}_{1 f} \gamma^\nu \left ( \gamma(m_1 n-q) + m_1 \right ) \gamma^\kappa \left ( \gamma(m_1 n-q-r) + m_1 \right ) \gamma^{\mu_1} }{ \left (q^2-2 qn \, m_1 \right )^2 ((q+r)^2-2 (q+r) n m_1)^2 } \crr
&\hspace{-0.4cm}& \times \frac{   \left ( \gamma(m_1 n-q-r+s) + m_1 \right ) \gamma^{\mu_2}  \left ( \gamma(m_1 n-q-r) + m_1 \right ) \gamma_\kappa \left ( \gamma(m_1 n-q) + m_1 \right ) \gamma_\nu u_{1 i} }{ \left ( (q+r-s)^2-2 (q+r-s)n \, m_1 \right ) } \crr
&\hspace{-0.4cm}& \times \frac{  \overline{u}_{2 f} \gamma_{\mu_1} \left ( \gamma(m_2 n-s)+m_2 \right ) \gamma_{\mu_2} u_{2 i} }{ \left (s^2-2 sn \, m_2 \right ) \left (q^2 \right ) \left ( r^2 \right ) \left ( s^2 \right )^2 }  .
\eea
\end{widetext}

We combine the traces using the identities \cite{Czarnecki1999}
\bse \bea
u_1 u^T_2 C &=& \frac{1}{\sqrt 8} (\gamma n + 1 ) \begin{Bmatrix} \gamma_5 \cr -\gamma \epsilon_m \end{Bmatrix} \, , \\
C^{-1} \overline{u}^T_2 \overline{u}_1 &=& \frac{1}{\sqrt 8} \begin{Bmatrix} \gamma_5 \cr \gamma \epsilon_m^* \end{Bmatrix} \left ( \gamma n + 1 \right ) \, ,
\eea \ese
which are valid for $\begin{Bmatrix} {\rm spin\!-\!0} \\ {\rm spin\!-\!1} \end{Bmatrix}$ configurations of the constituent particles.  Here $C$ is the charge conjugation matrix, which satisfies $C^{-1} \gamma_\mu^{\rm T} C = - \gamma_\mu$, $\gamma_5$ is a matrix that anticommutes with the other gamma matrices $\gamma_5 \gamma_\mu \gamma_5 = - \gamma_\mu$ and satisfies $\gamma_5^2=1$, and $\ep_m = (0 , \hat \ep_m)$ is a polarization vector.  We compute traces in $d$ dimensions using the standard Dirac anticommutation relations for gamma matrices, $\gamma_\mu \gamma^\mu=d$, and ${\rm tr}[1]=4$.  (There are no subtleties associated with $\gamma_5$ since it always occurs in pairs.)  One can rearrange terms
to write
\bea
\bar u_{1f} X u_{1i} \times \bar u_{2f} Y u_{2i} &=& \bar u_{1f} X u_{1i} \times u^{\rm T}_{2i} C \tilde Y C^{-1} \bar u^{\rm T}_{2f} \crr
&\phantom{=}& \hspace{-3.0cm} = \frac{1}{8} {\rm tr} \left [ X (\gamma n+1) \begin{Bmatrix} \gamma_5 \tilde Y \gamma_5 \cr -\gamma \epsilon_m \tilde Y \gamma \epsilon^*_m \end{Bmatrix} (\gamma n + 1 ) \right ] \, ,
\eea
where $\tilde Y = C^{-1} Y^{\rm T} C$.  For the spin-1 term, we average over the spin directions and use
\bea
\gamma \epsilon_m \otimes \gamma \epsilon^*_m 
&\rightarrow& \gamma^\alpha \otimes \gamma^\beta \frac{1}{d-1} \sum_m (\ep_m)_\alpha (\ep^*_m)_\beta \crr
&=& \gamma^\alpha \otimes \gamma^\beta \frac{1}{d-1} \left ( n_\alpha n_\beta - g_{\alpha \beta} \right ) \, .
\eea
The amplitude $\M_{\rm A1}$ becomes
\bea
{\M}_{\rm A1} &=& i (q_1^2)^2 (q_1 q_2)^2 \int \dbar^d \! q \, \dbar^d \! r \, \dbar^d \! s \, \frac{N_{\rm A1}}{ D_{\rm A1}} \crr
&=& \frac{\alpha^2 (Z \alpha)^2}{(4\pi)^2} \bar \mu^{2\epsilon} \mu^{6\epsilon} \Phi^3 \int \dbar^d \! q \, \dbar^d \! r \, \dbar^d \! s \, \frac{N_{\rm A1}}{ D_{\rm A1} } \, .
\eea
We have used $q_1 q_2 = -4 \pi Z \alpha \bar \mu^{2 \epsilon}$ for the product of the charges, and have included a mass parameter $\bar \mu^{2 \ep}$ that is necessary to keep the dimensions consistent in $d$ dimensions.  It is convenient to introduce an additional mass parameter $\mu$ and a factor $\Phi \equiv (-i) (4\pi)^{2-\epsilon} e^{\epsilon \gamma_E}$, defined so that
$\bar \mu^{2\epsilon} = \left ( \frac{\mu^2 e^{\gamma_E}}{4\pi} \right )^\epsilon = \mu^{2\ep} \frac{i \Phi}{(4\pi)^2}$.
We include one factor of $\Phi$ with each loop integral so that the results of the integrals, when expanded in the dimensional regularization parameter $\ep$, have as simple a form as possible without extraneous factors of $\gamma_E$ or $\ln(4 \pi)$.  The numerator in ${\cal M}_{\rm A1}$ is
\begin{widetext}
\bea
N_{\rm A1} &=& \frac{1}{8} {\rm tr} \Big [ \gamma^\nu \left ( \gamma(m_1 n-q) + m_1 \right ) \gamma^\kappa \left ( \gamma(m_1 n-q-r)+m_1 \right ) \gamma^{\mu_1}  \left ( \gamma(m_1 n-q-r+s) + m_1 \right ) \gamma^{\mu_2}  \crr
&\hbox{}& \hspace{-1.0cm} \times  \left ( \gamma(m_1 n-q-r) + m_1 \right ) \gamma_\kappa \left ( \gamma(m_1 n-q) + m_1 \right ) \gamma_\nu 
\left ( \gamma n + 1 \right ) \begin{Bmatrix} \gamma_{\mu_2} ( \gamma(m_2 n-s) + m_2 ) \gamma_{\mu_1}   \cr \gamma \epsilon_m \gamma_{\mu_2} (\gamma(m_2 n-s) - m_2) \gamma_{\mu_1}  \gamma \epsilon_m^*  \end{Bmatrix} \left ( \gamma n + 1 \right ) \Big ] \, ,
\eea
and the denominator is
\bea
D_{\rm A1} \! \! &=& \! \! (-q^2)(-r^2)(-s^2)^2 (-q^2 \! + \! 2 m_1 qn)^2 (-s^2 \! + \! 2 m_2 sn) [-(q+r)^2 \! + \! 2 m_1 (q+r)n]^2 [-(q+r-s)^2 \! + \! 2 m_1 (q+r-s)n ] \, . \crr
\eea
\end{widetext}
For the crossed graph everything is the same except that $\mu_1 \leftrightarrow \mu_2$ and $s \rightarrow -s$ in the bracketed term for the muon line, which becomes
\be
\begin{Bmatrix} \gamma_{\mu_1} ( \gamma(m_2 n+s) + m_2 ) \gamma_{\mu_2}   \cr \gamma \epsilon_m \gamma_{\mu_1} (\gamma(m_2 n+s) - m_2) \gamma_{\mu_2}  \gamma \epsilon_m^*  \end{Bmatrix} \, .
\ee
Our calculations were done using Mathematica \cite{mma143}.  After first evaluating the traces with FeynCalc \cite{FeynCalc1,FeynCalc2,FeynCalc3}, we used the integration by parts (IBP) identities \cite{Tkachov1981,Chetyrkin1981} and the program FIRE \cite{Smirnov2015} to reduce the expressions to a sum of 139 master integrals.  Many of the three-loop master integrals, having up to eight denominators, were difficult to evaluate analytically, so we elected to pursue a numerical evaluation for specified values of $x=m_1/m_2$ using the AMFlow program \cite{Liu2023}.  This program is able to evaluate dimensionally regularized Feynman integrals order by order in $\ep$, up to a specified maximum order, to a specified numerical precision.  The program generates sets of differential equations satisfied by the required integral and companion integrals using an auxiliary mass parameter as the independent variable.  The program solves the differential equations by a power series approach, using analytic continuation to move from a point where the boundary conditions are convenient to the physical point specified by the user.  The boundary conditions are themselves Feynman integrals that are dealt with by the program recursively and automatically.  The coefficients we found that multiplied the master integrals diverged, at most, as $1/\ep^4$ for $\ep$ approaching zero, so we had AMFlow evaluate the master integrals up to order $\ep^4$, for eight total orders in $\ep$ from $\ep^{-3}$ to $\ep^4$.  We specified 42 or more significant figures of precision from AMFlow to overcome any severe cancellation between terms with the concomitant loss of precision.

\begin{figure}[t!]
\begin{center}
\begin{minipage}{0.99\linewidth}
\begin{center}
\includegraphics[width=0.99\linewidth]{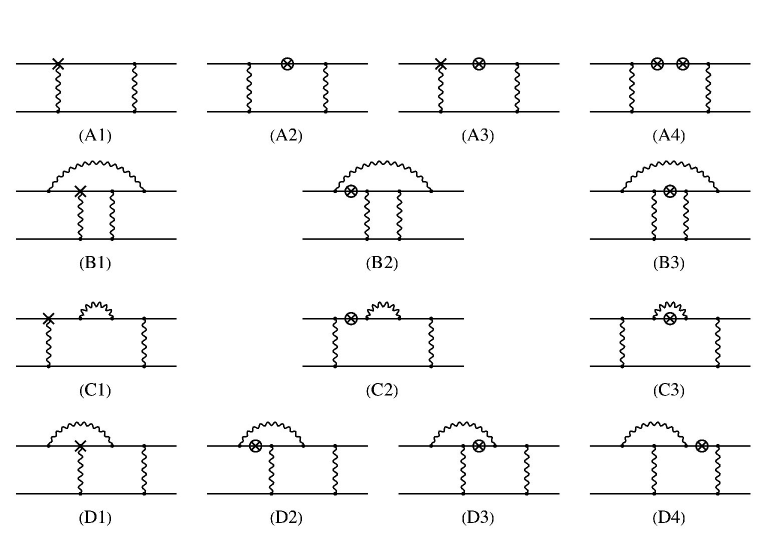}
\end{center}
\caption{\label{fig2}
Counterterm diagrams for the radiative-recoil correction of order $\alpha^2 (Z \alpha)^5$.  Graphs with crossed exchange photons must also be included.  Symmetry factors of two are included for diagrams (B2), (C2), and (D1)-(D4).}
\end{minipage}
\end{center}
\end{figure}

After evaluating the master integrals, the complete contribution of the diagrams of Fig.~\ref{fig1} was obtained for the  spin-0 and spin-1 states individually.  The total result contained $1/\ep^3$ divergences that required renormalization.  We used renormalized perturbation theory \cite{IandZ1980,Kaku1993,Weinberg1995,PeskinandSchroeder1995}, which contains vertex and self-mass counterterms, to arrive at the complete set of required graphs that, together, produce the finite renormalized scattering amplitudes.  Appropriate Feynman rules can be found in Ref.~\cite{Adkins2026} for the counterterm graphs of Fig.~\ref{fig2}.

All calculations using FeynCalc and FIRE were independently confirmed by multiple authors.  The AMFlow integrals were confirmed by comparison against analytic results for cases where the analytic results were available.

\begin{figure}[t!]
\begin{center}
\begin{minipage}{0.99\linewidth}
\begin{center}
\includegraphics[width=0.99\linewidth]{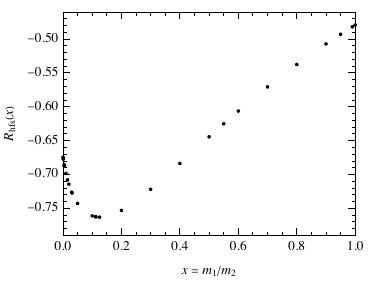}
\end{center}
\caption{\label{fig3}
Plot of $R_{\rm hfs}(x)$ showing the recoil dependence of the hyperfine structure contribution of order $\alpha^2 (Z \alpha)^5$.  Uncertainties are much too small to be visible.}
\end{minipage}
\end{center}
\end{figure}

{\em Results.---} We present our results as functions of the mass ratio $x$ with appropriate factors.  For the hyperfine splitting we have
\be
\Delta E_{\rm hfs} = \frac{\alpha^2 (Z \alpha)}{\pi n^3} E_F R_{\rm hfs}(x) \, ,
\ee
where the Fermi energy is
\be
E_F = \frac{8}{3} m_1 (Z \alpha)^4 \left ( \frac{m_1}{m_2} \right ) \left ( \frac{m_r}{m_1} \right )^3 \, .
\ee
For the average energy we have
\be
\Delta E_{\rm avg} = \frac{m_1 \alpha^2 (Z \alpha)^5}{\pi n^3} \left ( \frac{m_r}{m_1} \right )^3 R_{\rm avg}(x) \, .
\ee
We assume $m_1 \le m_2$.  Particle one is typically the electron, while particle two might be the proton, positive muon, or positron.  Plots of $R_{\rm hfs}(x)$ and $R_{\rm avg}(s)$ are shown in Figs.~\ref{fig3} and \ref{fig4}.

The recoil behavior of the hyperfine structure energy shift is shown in Fig.~\ref{fig3}.  The data for this plot is available in the Ancillary Material \cite{SupplementalMaterial} that accompanies this article.  The data points are quite precise, with ones near the center of the plot having up to 32 significant figures.  We can compare the no-recoil limit of our result, $-0.673259$
 (see Table I), against prior work.  Kinoshita and Nio obtained an early estimate of $R_{\rm hfs}(0)$ of -0.63(4) in 1994 \cite{Kinoshita1994}.  Eides and Shelyuto obtained -0.6726(4) in 1995 \cite{Eides1995}, and Kinoshita and Nio improved their earlier estimate in 1996, finding the value -0.6764(79) \cite{Kinoshita1996}.  A significantly more precise value, -0.67326(2), can be extracted from the work of Mond{\' e}jar, Piclum, and Czarnecki, published in 2010 \cite{Mondejar2010}.  Our no-recoil limit is consistent with these prior results.  All of these calculations, including ours, are based on the evaluation of a two-body scattering amplitude at threshold.  The Kinoshita-Nio calculation was done using Feynman gauge and a non-zero photon mass $\lambda$ for IR regulation, with the uncertainty dominated by the challenge of taking the $\lambda \rightarrow 0$ limit.  Eides and Shelyuto used the Fried-Yennie gauge, which has better IR behavior that allowed for higher precision.  Mond{\'e{har {\em et al.} used the integration by parts identities to express their result in terms of a few master integrals, which were evaluated numerically.

\begin{figure}[t!]
\begin{center}
\begin{minipage}{0.99\linewidth}
\begin{center}
\includegraphics[width=0.99\linewidth]{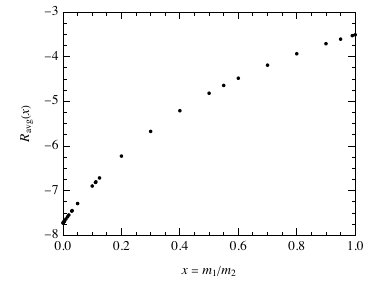}
\end{center}
\caption{\label{fig4}
Plot of $R_{\rm avg}(x)$ showing the recoil dependence of the spin-average contribution of order $\alpha^2 (Z \alpha)^5$.  Uncertainties are much too small to be visible.}

\end{minipage}
\end{center}
\end{figure}

\begin{table}
\begin{center}
\caption{\label{results} Order $\alpha^2 (Z \alpha)^5$ radiative-recoil corrections for various physical systems composed of spin-$\frac{1}{2}$ particles.  Our results are given for both the hyperfine splitting [$R_{\rm hfs}(x)$] and average [$R_{\rm avg}(x)$] correction functions.  Uncertainties are less than one in the last digit.}
\begin{ruledtabular}
\begin{tabular}{cccc}
System & $x=m_1/m_2$ & $R_{\rm hfs}(x)$ & $R_{\rm avg}(x)$  \\
\hline\noalign{\smallskip}
no recoil & 0 & -0.673259 & -7.723827 \\
hydrogen & $m_e/m_p \approx 0.000\,54$ & -0.675641 & -7.718773 \\
muonium & $m_e/m_\mu \approx 0.0048$ & -0.687745 & -7.679183 \\
muonic hydrogen & $m_\mu/m_p \approx 0.11$ & -0.762530 & -6.801971 \\
positronium & 1 & -0.479248 & -3.510425 \\
\end{tabular}
\end{ruledtabular}
\end{center}
\end{table}

The function $R_{\rm avg}(x)$ for the average energy correction is presented in Fig.~\ref{fig4}.  The no-recoil limit of this plot is -7.723827 as shown in Table~I.  The first  calculation of this no-recoil correction was done by Pachucki \cite{Pachucki1994}.  His result was -7.61(16), done using a hybrid method where some diagrams (a gauge-invariant set) were calculated in Fried-Yennie gauge, while the others were done using Feynman gauge and a mixed analytic/numeric approach.  Eides and Shelyuto \cite{Eides1995} achieved a higher precision result -7.724(1) using the Fried-Yennie gauge.  Dowling {\em et al.} \cite{Dowling2010} found the high-precision result -7.72381(4) by utilizing the integration by parts identities and a numerical evaluation of the resulting master integrals.

{\em Conclusions.---} We calculated the radiative-recoil corrections of order $\alpha^2 (Z \alpha)^5$ to the hyperfine splitting and average energies of $S$ states of two-body QED bound systems for all values of the mass ratio.  Our muonium ground state hyperfine result is
\be
\Delta E_{\rm hfs} = \frac{\alpha^2 (Z \alpha)}{\pi} E_F  R_{\rm hfs}(x_{\rm M}) = -378.886(1) \, {\rm Hz} \, ,
\ee
where $x_{\rm M} = m_e/m_\mu$.  This corrects the no-recoil estimate of Mond{\'e}jar {\em et al.} \cite{Mondejar2010} by $-7.98(1) \, {\rm Hz}$, where the no-recoil value was found using $R_{\rm hfs}(0) \approx -0.67326(2)$.  The $-7.98(1) \, {\rm Hz}$ recoil correction is relevant given the expected precision of $\sim \! 10 \, {\rm Hz}$ of the MuSEUM experiment \cite{Iwai2023,Strasser2025} and will reduce the theoretical uncertainty of $\pm 70\,{\rm Hz}$ cited by Eides \cite{Eides2019,Eides2026} or $\pm 85\,{\rm Hz}$ cited by Mohr {\it et al.} \cite{Mohr2025}.  For positronium the $\alpha^2 (Z \alpha)^5$ correction is
\be
\Delta E_{\rm hfs} = 2 \times \frac{\alpha^2 (Z \alpha)}{\pi} E_F  R_{\rm hfs}(1) = -13.85 \, {\rm kHz} \, .
\ee
(The contribution was doubled to account for two-loop radiative corrections on the positron line.)

We also have new results for the muonium and positronium 1S-2S transition energies
\be
\Delta E_{\rm avg}(1S-2S) = - \frac{7 m_e \alpha^7}{8 \pi} \left ( \frac{m_r}{m_e} \right )^3 R_{\rm avg}(x) \, .
\ee
For muonium, our radiative-recoil result is $287.02445(4) \, {\rm kHz}$, which corrects the no-recoil result of Dowling {\rm et al.} \cite{Dowling2010} by $-1.668(2) \, {\rm kHz}$ (to be compared with the $\sim 10\,{\rm kHz}$ goal of the Mu-MASS experiment \cite{Cortinovis2023}), while for positronium the contribution at this order is $33.28 \, {\rm kHz}$.

We have given high-precision results for the order $\alpha^2 (Z \alpha)^5$ radiative-recoil corrections for all values of the mass ratio.  These results are important given the expected uncertainties of ongoing experiments, and they help to complete the calculation of all corrections to these systems of order $\alpha^7$.  In addition, our calculation demonstrates the usefulness of the program AMFlow \cite{Liu2023}, which allows for the convenient evaluation of dimensionally regularized multiloop integrals to high precision.

{\em Acknowledgements.---} We acknowledge the assistance of Lecheng Ni in the early stages of this project.  This work was completed in part using the F\&M research cluster, supported by F\&M’s Research Computing team and NSF CC* Cyber-infrastructure award 1925192.  Crucial support was provided by the National Science Foundation grant No.~PHY-2308792 and by the National Institute
of Standards and Technology grant  60NANB23D230.

{\em Data Availability.---} The data that supports the findings of this article are available in the accompanying Ancillary Materials.

\end{document}